\documentstyle[11pt]{article}

\begin{document}
\thispagestyle{empty}
\begin{center}
\rightline{IC/2002/78}
\vspace{2cm}
{\bf  HYBRID INFLATION IN SUPERGRAVITY WITHOUT\\ INFLATON 
SUPERPOTENTIAL}\\
\vspace{2cm}
H. Boutaleb-J\\
{\it Laboratoire de Physique Th\'eorique, Facult\'e des 
Sciences,\\
BP 1014, Agdal, Rabat, Morocco,}\\[1.5em]
A. Chafik\\ 
{\it Laboratoire de Physique Th\'eorique, Facult\'e des 
Sciences,\\
BP 1014, Agdal, Rabat, Morocco,}\\[1.5em]
A.L. Marrakchi\\
{\it Laboratoire de Physique Th\'eorique, Facult\'e des 
Sciences,\\
BP 1014, Agdal, Rabat, Morocco\\
and\\
D\'epartement de Physique, Facult\'e des Sciences,\\
BP 1796, Atlas, F\`es, Morocco.} \\[1.5em]
\end{center}
\vfill

\newpage

\centerline{\bf Abstract}
\baselineskip=20pt
\bigskip

We propose a new realisation of hybrid inflation in supergravity
where the inflaton field does not appear in the superpotential but
contributes only through the K\"ahler potential. The scalar potential 
derived
from an $R$-invariant superpotential has the same form as that of the 
Linde's
original version. The correct magnitude of the density perturbations 
amplitude
is found without any fine-tuning of the coupling parameter in the
superpotential for an acceptable value of the fundamental energy scale 
of the
theory. The $\eta-$problem was also resolved in this model.\\ \\
PACS number(s): 98.80.Cq

\newpage

\section{Introduction}

It is now believed that it is necessary to invoke an inflationary era 
\cite{key1} in order to give a consistent description of the early 
Universe. Indeed, the standard hot big-bang model presents some 
conceptual 
problems related to the requirement of unnatural initial conditions, in 
particular the horizon and flatness problems. These problems have been 
resolved by assuming the existence of a sufficiently long period of 
fast inflation during which the energy density of the Universe is 
dominated by a non-vanishing vacuum energy. In such a scenario we can 
also 
explain the origin of the density perturbations which are responsible 
for 
the observed temperature anisotropies in the cosmic microwave 
background 
radiation (cmbr) and the large-scale structure of the observable 
Universe.

Supersymmetry which was initially motivated by the gauge hierarchy 
problem \cite{key2} seems to have interesting cosmological implications 
as 
well. On one side, supersymmetric theories often have non compact flat 
vacuum directions which remain flat to all order of perturbation theory 
\cite{key3}. They are then good candidates for inflation since a long 
period of inflation requires a sufficiently flat potential. The 
combination of inflation and supersymmetry naturally leads to 
consider inflation in supergravity. However, the single-field models of
inflation present a common problem of naturalness. Indeed, to 
accomplish a
successful inflationary scenario the corresponding constraints on the 
potential
of the inflaton impose an unrealistic fine tuning of the parameters of 
the
relevant theory of particle physics. This problem can be avoided in the 
hybrid
inflation model proposed by Linde \cite{key4}. 

The hybrid inflation model is naturally realised in supersymmetric 
theories and
its relevance to SUSY has been extensively investigated \cite{key5}. 
The non
zero vacuum energy density during inflation can either be due to the 
vev of a
$F$-term or that of a $D$-term. The scalar potential has two minima: a 
local
one of value of the inflaton $S$ greater than some critical value $S_c$ 
(with
a vanishing noninflaton field $\phi$), and a global supersymmetric one 
at
$S=0$. When $S\gg S_c$ the universe is dominated by a non
vanishing vacuum energy, the slow rolling conditions are satisfied and
inflation takes place until $S=S_c$ when a phase transition occurs 
causing
the end of inflation.

The naturalness issue has been a great challenge for physicists. In 
this paper we want to contribute to the resolution of this problem in 
the 
case of a model of inflation in supergravity based on an $R-$invariant 
superpotential. Our approach consists of constructing a hybrid 
inflation 
model from the same superpotential as that of the 
model of reference \cite{key6} where the authors have 
obtained a flat potential as usually required for a successful 
inflation, but 
at the price of a very small coupling parameter. We have introduced a 
second scalar field that does not contribute to the superpotential but 
appears
only in the K\"ahler potential. The result is a hybrid inflation model 
where
the  non-vanishing energy is provided by the inflaton potential.

\section{The initial model}
In this section we give a brief review of the model presented in the 
reference \cite{key6}. This model is based on the discrete $Z_n$-R 
invariance  which is given by the transformation:   
\begin{equation}
\phi(x,\theta)\longrightarrow 
e^{-i\alpha_n}\phi(x,e^{i\alpha_n/2}\theta)
\end{equation}
on the inflaton field where $\alpha_n\,=\,\frac{2\pi k}{n}(k=\pm 1, \pm 
2, \cdot\cdot\cdot)$.

The general form of the superpotential and the K\"ahler potential which 
have the $Z_n-R$ invariance is given by:
\begin{equation}
W_n(\phi)=\phi\sum_{l=0}^{\infty}b_l\phi^{ln}
\end{equation}
\begin{equation}
K(\phi,\phi^*)=\sum_{m=1}^{\infty}a_m(\phi\phi^*)^m
\end{equation}

The K\"ahler potential is taken to be of the minimal form:
\begin{equation}
K(\phi,\phi^*)=\frac{\phi\phi^*}{M_p^2}
\end{equation}

The expression Eq.(2) is convergent only for $\vert\phi\vert\le 0$, 
this corresponds also to a model of new inflation where the inflaton 
field 
begins its evolution near the origin. So, in order to find the 
approximate form of $V(\phi)$ near the origin $\phi\sim 0$, one can 
take only the 
terms:
\[
b_0\,=\,\lambda v^2 \:\:\:;\:\:\: 
b_1\,=\,\frac{\lambda}{v^{n-2}}\frac{1}{n+1}
\]
the superpotential is then written as:
\begin{equation}
W(\phi)=\biggl(\frac{\lambda}{v^{n-2}}\biggr)\biggl(v^n\phi-\frac{1}{n+1}\phi^{n+1}\biggr)+\cdot\cdot\cdot
\end{equation}
where $\lambda$ is a dimensionless coupling constant and $v$ is a 
constant of  dimension one in mass unit\footnote{In fact, $v$ is a 
scale of 
the condensation of a superfield coupled to the inflaton. This 
condensation breaks a $U(1)$ symmetry down to the discrete $Z_n -R$ 
symmetry.}. 
In the above equation ... represent higher power part  ($\phi^{kn+1}$ 
with $k\geq 2$).\\

In the minimal $N=1$ supergravity a scalar potential $V$ is written as 
\cite{key2}

\begin{eqnarray}
V(\phi)\, &=& 
\,e^{\vert\phi\vert^2/M_p^2}\biggl\lbrace\biggl\vert\frac{\partial 
W_n}{\partial\phi}\,+\,\phi^*\frac{W_n}{M_p^2}\biggr\vert^2\, 
-\,\frac{3\vert W_n\vert^2}{M_p^2}\biggr\rbrace \nonumber \\
& + & D-term
\end{eqnarray}
where $M_p$ is the reduced Planck mass $M_p=m_p /\sqrt{8\pi}=2.4\times 
10^{18} GeV$.\\

With the expressions (4) and (5), and the equation (6), the $Z_n -R$ 
invariant scalar potential becomes 
\begin{equation}
V(\phi)\,=\,\biggl(\frac{\lambda}{v^{n-2}}\biggr)^2\Biggl(v^{2n}+\frac{1}{2}v^{2n}\biggl(\frac{\vert\phi\vert^2}{M^2}\biggr)^2-v^n\bigl(\phi^n+\phi^{*n}\bigr)\Biggr)
\end{equation}

Since $v\ll M_p$ the $\phi^2/M_p^2$ term can be neglected for $n\geq 3$ 
which gives a very flat region in the inflaton potential near $\phi=0$. 
Identifying the inflaton field with the real component of $\phi$ 
($\varphi=\sqrt{2}Re\phi$), the relevant potential is now
\begin{equation}
V(\varphi)\simeq\tilde{\lambda}^2\tilde{v}^4\biggl[1-2\biggl(\frac{\varphi}{\tilde{v}}\biggr)^n\biggr]
\end{equation}
with $\tilde{\lambda}=\frac{1}{2}\lambda$ and $\tilde{v}=\sqrt{2}v$.

During the slow-rolling phase the inflationary dynamics is described by 
the equation of motion:
\begin{equation}
\dot{\varphi}\, \simeq 
\,\frac{2n\tilde{\lambda}M_p}{\sqrt{3}\tilde{v}^{n-1}}\varphi^{n-2}
\end{equation}

The slow-rolling regime ends at:
\begin{equation}
\varphi_f^{n-2}\,\simeq\,\frac{1}{6n(n-1)}\frac{\tilde{v}^n}{M_p^2}
\end{equation}

From the constraint imposed by the observed anisotropies on the 
amplitude of the density perturbations one deduces the equation:
\begin{equation}
\frac{\tilde{\lambda}\tilde{v}^2}{10\sqrt{3}\pi 
nM_p^3}\bigg\lbrace\frac{\tilde{v}^2}{2Nn(n-2)M_p^2}\bigg\rbrace^{\frac{1-n}{n-2}}\,\sim\, 
2\times 10^{-5}
\end{equation}
where $N$ is the total number of e-foldings of the inflationary 
phase.\\

The gravitino mass $m_{3/2}$ is given by:
\begin{equation}
m_{3/2}\,\simeq\, 
\frac{n}{\sqrt{2}(n+1)}\tilde{\lambda}\tilde{v}\biggl(\frac{\tilde{v}}{M_p}\biggr)^2
\end{equation}

From Eqs.(11) and (12) one can determine $\lambda$ and $v$ for a given 
sets of $N$ and $m_{3/2}$. To illustrate the results we choose, for 
example, the case $v\sim 10^{15}$ GeV, $\lambda$ is then:
\begin{equation}
\lambda\,\sim\, 5\times 10^{-7}
\end{equation}

\section{The hybrid inflation model}
The authors have clearly succeeded to obtain a sufficiently flat 
potential for a successful inflationary model (as we can see from 
Eq.(8)). 
However, the price to pay was the fine-tuning of the coupling parameter 
[Eq.(13)] as was always the case in one field models of inflation, in
particular the fine tuning was the main problem of the new inflation 
scenario
implemented in this model. Indeed,  small parameters are inevitable in 
order to
have sufficient inflation  and the correct magnitude of density 
fluctuations.
The resolution of  this problem was one of the initial motivations of 
the
hybrid inflation model proposed by Linde \cite{key4} and studied by 
Copeland
{\it et al.}~\cite{key7} where two scalar fields are relevant. 

In the same way we try to construct a hybrid inflation model in 
supergravity.  Our model is based on the superpotential given by Eq.(5) 
with 
the introduction of a second scalar field in such a
way to preserve the initial $R-$invariance. The second field we will 
introduce
in this model contributes only through the K\"ahler potential.  The 
G-singlet
fields which do not contribute to the superpotential have been 
considered in
reference \cite{key8} to generate the mass-term of the inflaton field. 

In this letter we show that such a field can play a more important role 
than
simply contribute to the mass of the inflaton: it could even play the 
role of
the inflaton. A simple realisation of such a model is provided by the
superpotential [Eq.(5)] and the K\"ahler potential:
\begin{equation}
K\; =\;\frac{\phi^*\phi}{M_p^2}\; +\;\alpha\frac{S^* S}{M_p^2}
\end{equation} 
where $\alpha$ is a small parameter ($\alpha\ll 1$).\\ 

This form of the K\"ahler potential clearly respects the initial 
discrete
$R-$invariance Eq.(3) for the two fields. \\

The scalar potential is given by:
\begin{eqnarray}
V(\phi , S)\; & = &\; \mbox{exp}\Biggl(\bigl(\vert\phi\vert^2 
+\alpha\vert
S\vert^2\bigr)
/M_p^2\Biggr)\biggl(\frac{\lambda}{v^{n-2}}\biggr)^2\Biggl\{\biggl\vert 
v^n
-\phi^n +\frac{\phi^*}{M_p^2}\biggl(v^n\phi -
\frac{1}{n+1}\phi^{n+1}\biggr)\biggr\vert^2 \nonumber \\  &\;\;& +\;\;
\biggl\vert \alpha\frac{S^*}{M_p^2} \biggl(v^n\phi
-\frac{1}{n+1}\phi^{n+1}\biggr)\biggr\vert^2 \;\; -\;\;
\frac{3}{M_p^2}\biggl\vert v^n\phi 
-\frac{1}{n+1}\phi^{n+1}\biggr\vert^2
\Biggr\} 
\end{eqnarray}

By taking the leading terms in the exponential factor to be : $1 + 
\alpha
\vert S\vert^2 /M_p^2$ the scalar potential takes the form:

\begin{eqnarray}
V(\phi ,S)\;\;& = &\;\; \biggl(\frac{\lambda}{v^{n-2}}\biggr)^2\Biggl\{
P_1(\phi^m)\; -\; 2\frac{\vert\phi\vert^2}{M_p^2}P_2(\phi^n)  \nonumber 
\\ 
&\;\;& +\;\; \frac{\vert\phi\vert^4}{M_p^4}P_2(\phi^n) \; 
+\;\alpha\frac{\vert
S\vert^2}{M_p^2}P_3(\phi^n) \; +\;
\alpha\frac{\vert S\vert^2\vert\phi\vert^2}{M_p^4}P_2(\phi^n) \nonumber 
\\
&\;\;& \;\; +\;\;
\mbox{higher powers of} \vert\phi\vert^n\Biggr\}  
\end{eqnarray}
where $P_i(\phi^n)= v^{2n} - c(n)(\phi^n+\phi^{*n})$ ($i=1,2,2,4$) and 
$c(n)$
is a function of n.

The higher powers of $(\phi/v)^n$ do not contribute to the dynamics of 
the
inflaton field and can be ignored. By appropriate transformation we can 
bring
$S$ and $\phi$ to the real axis:
\begin{equation}
S=\frac{\sigma}{\sqrt{2}}\;\;\; ; \;\;\; \phi=\frac{\varphi}{\sqrt{2}}
\end{equation}

The relevant scalar potential is then:
\begin{equation}
V(\sigma ,\varphi)\;\;=\;\; \lambda^2v^4\Biggl[\biggl(1 -
\frac{\varphi^2}{2 M_p^2}\biggr)^2\; +\; \frac{\sigma^2\varphi^2}{4
M_P^4}\; +\; \alpha\frac{\sigma^2}{2 M_p^2}\Biggr]
\end{equation}
which can be written in the famous form:
\begin{equation}
V(\sigma , \psi)\;\; =\;\; \frac{1}{4}\lambda'\bigl(M^2 
-\psi^2\bigr)^2\; +\;
\frac{g^2}{2}\sigma^2\psi^2\; +\;\frac{1}{2}m_{\sigma}^2\sigma^2
\end{equation}
by means of the following rescaling:
\begin{eqnarray}
M\; &\equiv &\; v \\
\lambda'\; &\equiv &\; (2\lambda)^2 \\
g^2\; &\equiv &\; \frac{\lambda^2v^2}{M_p^2} \\
\psi\; &\equiv &\; \varphi\frac{v}{\sqrt{2}M_p} \\
m_{\sigma}^2\; &\equiv &\; \alpha\frac{\lambda^2 v^4}{M_p^2}
\end{eqnarray}

A detailed investigation of such a model has been presented by Copeland 
{\it
et al.}~\cite{key7}. For values of $\sigma$ larger than:
\begin{equation}
\sigma_c\;\; =\;\; 2M_p
\end{equation}
the minimum of $V$ is at $\varphi=0$, the energy density of the 
universe is
dominated by the potential energy of the scalar field $\sigma$:
\begin{equation}
V(\sigma)\;\; =\;\;\frac{1}{4}\lambda'M^4\; +\; 
\frac{1}{2}m_{\sigma}^2\sigma^2
\end{equation}

Inflation ends when $\sigma$ falls below $\sigma_c$ and the fields 
rapidly
adjust to their true vacuum values ($\varphi = M_p$ and $\sigma=0$) 
with a
vanishing cosmological constant $V=0$. \footnote{The vanishing 
cosmological
constant in the true vacuum state is a characteristic feature of hybrid
inflation. This is in fact an advantage of our model since in reference
\cite{key6} the cosmological constant was negative in the global 
minimum of
the potential, the authors have introduced a $U(1)$ gauge multiplet in 
the
hidden sector and added a Fayet-Iliopoulos $D$-term so as to cancel the 
non
vanishing cosmological constant, and this looks very artificial.} 
However,
inflation can end before $\sigma$ reaches its critical value, when the
potential becomes too steep to maintain the slow rolling. This happens 
when
the slow rolling conditions \cite{key9}:
\begin{eqnarray}
\epsilon(\sigma)\;\; &\equiv&
\;\;\frac{M_p^2}{2}\biggl(\frac{V'(\sigma)}{V(\sigma)}\biggr)^2\;\;\ll 
1 \\
\eta(\sigma)\;\; &\equiv& \;\; M_p^2\frac{V''(\sigma)}{V(\sigma)}\;\; 
\ll
1\;\;\; ,
\end{eqnarray}
cease to be valid. The corresponding value of $\sigma$ 
($\epsilon(\sigma_e)=1$)
is given by \cite{key5} :

\begin{equation}
\sigma_e\;\; =\;\;\frac{M_p}{\sqrt{16\pi}}\biggl(1\;
+\;\sqrt{1-\frac{8\pi}{M_p^2}\frac{\lambda'M^4}{m_{\sigma^2}}}\;\;\biggr)
\end{equation}
which does not exist in our model. Hence, the end of inflation 
coincides with
$\sigma_c$. Note that the value $\sigma_c \simeq M_p$ corresponds also 
to the end of
inflation in the Linde's chaotic inflation \cite{key10} which recalls 
the fact
that hybrid inflation is initially a hybrid between chaotic inflation 
and
phase transition based models of inflation. Furthermore, it is clear 
that
during the inflationary phase the energy density of the universe is 
dominated
by the constant false vacuum energy term in Eq.(26).

The number $N$ of e-foldings of expansion which occur between two 
scalar field
values is given by the expression:
\begin{equation}
N(\sigma_1 ,\sigma_2)\;\; =\;\;
-\frac{1}{M_p^2}\int_{\sigma_1}^{\sigma_2}\frac{V(\sigma)}{V'(\sigma)}d\sigma
\end{equation}

In our approximation this gives:
\begin{equation}
N(\sigma_1 ,\sigma_2)\;\; =\;\;
-\frac{1}{2 M_p^2}\int_{\sigma_1}^{\sigma_2}\sigma d\sigma
\end{equation}

If we take $\sigma_2 = \sigma_c = 2 M_p$, the condition of sufficient
inflation $N\geq 70$ translates to the constraint:
\begin{equation}
\sigma_1 \geq 8 M_p
\end{equation}

This constraint justifies the choice of the parameter $\alpha$ as it 
was
introduced in Eq.(14).\\ 

 According to the analysis
of reference \cite{key7} the CMB constraint in this case is given by:
\begin{equation}
\frac{1}{M_p}\lambda M\;\; \leq\;\; 5\times 10^{-2}
\end{equation}
which translates in our model to the equation:
\begin{equation}
\lambda\biggl(\frac{v}{M_p}\biggr)^4\;\; \leq \;\;5.9\times 10^{-6}\\
\end{equation}

This can be achieved, for instance, by the choice : $\lambda\sim 
10^{-2}$ and
$v\sim 10^{16}$ GeV, which is very acceptable.

Now let us comment on the value $\sigma_c$ Eq.(25). The fact that 
$\sigma >
M_p$ during inflation is problematic if one considers the full 
expansion of
$K$ in the $S-$direction (Eq.(3)), since all the terms should be of the 
same
order of magnitude and have important contributions in the scalar 
potential.
Indeed, if we consider, for instance, the next term in Eq,(14) so that:
\begin{equation}
K\;\; =\;\;\frac{\phi^*\phi}{M_p^2}\; +\;\alpha\frac{S^* S}{M_p^2}\;
+\;\beta\frac{(S^* S)^2}{M_p^4}
\end{equation}
and demand that $\beta\ll\alpha$, we will have an additional term in 
the scalar
potential: 
\begin{equation}
\Delta V\;\; =\;\;\frac{\lambda''}{4}\sigma^4
\end{equation}
where:
\begin{equation}
\lambda''\;\; =\;\;\beta\frac{\lambda^2 v^4}{M_p^4}
\end{equation}

(Note that the value of $\sigma_c$ remains unchanged.)\\

This contribution to the scalar potential can be smaller than that of 
the
mass-term only if:
\begin{equation}
\sigma\;\; <\;\;\sqrt{\frac{2\alpha}{\beta}}M_p
\end{equation}
which contradicts Eq.(25) when $\beta\ll\alpha$. However, we can always 
make
an appropriate choice of the omitted terms in Eq.(3) of the K\"ahler
potential ($a_1 = \alpha$ and other $a_i = 0$), it is possible to 
arrange for
the potential Eq.(18).

However, the new term does not introduce any fine-tuning in the model. 
If we
assume that $\Delta V > 1/2 m_{\sigma}^2\sigma^2$, the potential during
inflation takes the form:
\begin{equation}
V(\sigma)\;\; =\;\;\frac{\lambda'}{4}\bigl( M^4\; +\; B\sigma^4\bigr)
\end{equation}
where:
\begin{equation}
B\;\; =\;\;\frac{\beta}{2 M_p^2}
\end{equation}

The investigation of such a potential has been the concern of reference
\cite{key11} where several mechanisms of inflation have been studied. 
The CMB
constraint in the case corresponding to our model gives:
\begin{equation}
\lambda^2\beta\ll 10^{-4}
\end{equation}

It is obvious that we can still have $\lambda\sim 10^{-2}$ for 
$\beta\ll 1$.
This proves that even with higher powers in the expression Eq.(3) for 
the
inflaton field there is no fine-tuning of the coupling parameter. 

This model provides a solution to another problem: the so-called
$\eta-$problem, which is a generic feature of minimal-supergravity 
models of
inflation (see for example \cite{key7}). In such models there is a 
contribution
of order $H^2$ to the mass square of the inflaton during inflation 
which
gives a contribution of order unity to the slow-rolling parameter 
$\eta$ and
then destroys the slow-rolling conditions. It is clear that in our 
model:
\begin{equation}
m_{\sigma}^2\ll H^2\simeq \frac{\lambda^2 v^4}{3 M_p^2}
\end{equation}

We have constructed a hybrid inflation model without inflaton 
superpotential,
the inflaton field contributes only through the K\"ahler potential. 
This model
presents the advantage of the original version of hybrid inflation 
model,
namely the resolution of the fine-tuning problem. Another generic 
problem of
the supergravity models which is the $\eta-$problem has also been 
overcome.
On the other hand, while in the initial model the cosmological constant 
in the
ground state was negative, in the present model it vanishes without 
need of any
artificial mechanism to cancel it. Finally, since the higher powers of 
the non
inflaton field $\vert\phi\vert^n$ do not contribute to the dynamics of 
the
inflaton field, they can be neglected. The model is then independent of
the integer $n$, and applies to all $Z_n -R$ symmetries without any 
restriction.

\section*{Acknowledgments}

A. Chafik thanks G. Dvali for helpful discussions and acknowledges the
Abdus Salam ICTP for hospitality.

\end{document}